\documentclass[twocolumn,showkeys,preprintnumbers,amsmath,amssymb]{revtex4}

\usepackage{graphicx}
\usepackage{dcolumn}
\usepackage{bm}
\usepackage{epsfig}

\begin{document}


\title{Hydrogen bond dynamics at the glass transition}

\author{U. Buchenau}
 \email{buchenau-juelich@t-online.de}
\affiliation{%
Forschungszentrum J\"ulich GmbH, J\"ulich Centre for Neutron Science (JCNS-1) and Institute for Complex Systems (ICS-1),  52425 J\"ulich, GERMANY}%

\date{October 31, 2021}

\begin{abstract}
The glass transition in hydrogen-bonded glass formers differs from the glass transition in other glass formers. The Eshelby rearrangements of the highly viscous flow are superimposed by strongly asymmetric hydrogen bond rupture processes, responsible for the excess wing. Their influence on the shear relaxation spectrum is strong in glycerol and close to zero in PPE, reflecting the strength of the hydrogen bond contribution to the high frequency shear modulus. A recent theory of the highly viscous flow enables a quantitative common description of the relaxation spectra in shear, linear as well as non-linear dielectrics, and heat capacity.
\end{abstract}

\maketitle 

At the glass transition \cite{cavagna,bb,royall,berthier}, hydrogen bonds \cite{oh1,oh2} have specific dynamics. At long times, in the monoalcohols \cite{mono}, the hydrogen bond decay is a separate Debye process, with a relaxation time much longer than the terminal shear relaxation time $\tau_c$, so the hydrogen bond connection memory survives the breakdown of rigidity, similar to the case of polymers \cite{sokolov}. 

At short times, two peculiarities seem to appear exclusively in hydrogen-bonded glass formers, namely strongly asymmetric double-well potentials in the recoverable part of the flow relaxation \cite{olsen,gainaru} and the excess wing at very short relaxation times \cite{schneider,gainaru,gabriel,pabst}. Both features have very recently been shown to be absent in several non-hydrogen-bonding glass formers by combining mechanical data in the glass phase at many different frequencies from the literature \cite{abs}.

The first clear evidence for strongly asymmetric double-well potentials in hydrogen-bonded substances appeared in an aging experiment \cite{olsen}. An average asymmetry of 3.8 $k_BT_g$ is needed to explain the intensity rise of the strong secondary relaxation peak dielectric signal in tripropylene glycol immediately after the initial temperature jump.

The second proof for a strong asymmetry is the strong temperature dependence $\exp(5T/T_g)$ of the excess wing measured in the glass phase of glycerol and other hydrogen bonded glass formers \cite{gainaru}, explainable in terms of the asymmetry $\Delta=5k_BT_g$, leading to the weakening factor $1/\cosh(\Delta/2k_BT)^2\approx\exp(5(T-1.52T_g)/T_g)$ for $T$ slightly below $T_g$.  

The breaking of hydrogen bonds has been intensely studied in liquid water, and in the water shell of biomolecules \cite{bagchi,laage}. In liquid water at room temperature, a hydrogen bond between two water molecules has two lifetimes, a short reversible one of 0.5 ps, after which it breaks, links to another water molecule, but then returns to its former state, and a longer irreversible one of 6.5 ps \cite{bagchi}. Obviously, this short time process is a rupture and re-formation of the hydrogen bond leading to a metastable energy minimum lying higher than the initial one, with a high back-jump probability, precisely the kind of process needed to understand the strongly asymmetric double-well potentials in the recoverable compliance of hydrogen-bonded glass formers.

There is a theoretical analysis \cite{bu2018a,bu2018b} of the reversible and irreversible shear transformation processes \cite{plazek,tnb,bero,roland} in the five-dimensional shear space in terms of asymmetric double-well potentials, with the asymmetry determined by the different shear misfits of the inner Eshelby domain \cite{eshelby} or shear transformation zone \cite{falk,johnson} in its two structural alternatives. One finds an Eshelby region lifetime $\tau_c=8\eta/G$ ($\eta$ viscosity, $G$ short time shear modulus), eight times longer than the Maxwell time. In terms of the barrier variable $v=\ln(\tau/\tau_c)$, one can describe the shear relaxation of simple glass formers without secondary relaxation peaks in terms of the Kohlrausch barrier density
\begin{equation} \label{lv}
	l(v)=0.4409\frac{8-4\beta}{3}\exp(\beta v),
\end{equation}
where $\beta$ is the Kohlrausch exponent \cite{bnap,albena} close to 1/2.

The idea is that the Kohlrausch barrier density extends without discontinuity or change of slope to barriers with $v>0$, which are irreversible transitions and are responsible for the viscous flow. Their flow contribution does not diverge, because the increase with $\exp(\beta v)$ is overcompensated by the rate factor $1/\tau=1/\exp(v)\tau_c$. In simple cases, metallic glasses and molecular glass formers without hydrogen bonds and with no secondary relaxation peak \cite{bu2018b}, eq. (\ref{lv}) works, and the whole shear relaxation is described by the three parameters $G$, $\eta$ and $\beta$.

If one has a secondary shear relaxation peak from changes of the shape or the orientation of the molecule, these shape or orientation changes do not contribute to the viscosity. The viscous flow requires an irreversible change of the molecular packing, while changes of the molecular shape or orientation are reversible by definition. In these cases, one has to add an appropriate gaussian distribution $l_G(v)$ to describe the secondary relaxation peak. But then, one finds \cite{bu2018b} that one has to increase the Kohlrausch barrier density of eq. (\ref{lv}) by the factor
\begin{equation}
	f_K=1+\int l_G(v)dv,
\end{equation}
which obviously means that now the irreversible transitions contain a fraction $(f_K-1)/f_K$ of shape or orientation changes in their shear contribution which does not contribute to the viscous flow, consistent with the concept that a high-barrier relaxation is composed of many low-barrier relaxations.

In hydrogen bonding substances, one finds $f_K>1$ even if there is no secondary relaxation peak. Their shear relaxation is described by the reversible barrier density
\begin{equation}
l_{rev}(v)=0.4409f_K\frac{8-4\beta}{3}\exp(\beta v+f_{exc}v^2)\exp(-\exp(v))
\end{equation}
($\exp(-\exp(v))=\exp(-\tau/\tau_c)$ lifetime cutoff). The small parameter $f_{exc}$ describes the excess wing at very short relaxation times. But its integrated intensity is not large enough to explain the fitted $f_K$-values. Obviously, a large part of the shear relaxation of the irreversible processes is due to hydrogen bond orientation changes, which do not contribute to the viscous flow.

Having defined $l_{rev}(v)$, one can calculate the complex shear compliance $J(\omega)$ from
\begin{equation}\label{jom}
	GJ(\omega)=1+\int_{-\infty}^\infty \frac{l_{rev}(v)dv}{1+i\omega\tau_c\exp(v)}-\frac{i}{\omega\tau_M},
\end{equation}
where $\tau_M=\eta/G=\tau_c/8$ is the Maxwell time, and invert it to get $G(\omega)$.

\begin{figure}[t]
\hspace{-0cm} \vspace{0cm} \epsfig{file=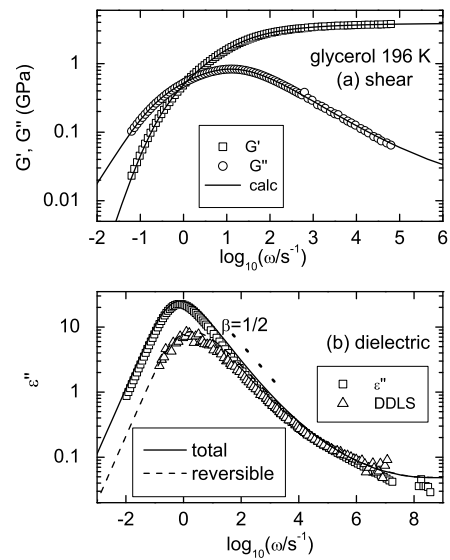,width=6 cm,angle=0} \vspace{0cm} \caption{(a) Measurement \cite{glynew} of $G(\omega)$ in glycerol, continuous lines calculated with the parameters in Table I (b) Fit of dielectric data \cite{roessler} at the same temperature with the parameters in Table II. The dashed line is calculated without the irreversible processes and happens to provide a good description of the depolarized dynamic light scattering data \cite{gabriel} (but see also Supplemental Material).}
\end{figure}

Fig. 1 (a) shows the fit of the shear relaxation data \cite{glynew} in glycerol at 196 K in terms of these equations, with the parameters compiled in Table I, demonstrating that the postulate of an additional slow mode \cite{glynew} is not the only way to understand these data. 

But then, one has to answer the question what the high value $f_K=5.26$ found in glycerol means. From the water evidence \cite{bagchi,laage}, one guesses that eighty percent of the shear fluctuations around $\tau_c$ arise from hydrogen bond breaking (reversible or irreversible), and only twenty percent from the separation of neighboring molecules. Of course, these two processes are not independent of each other; in order to separate two molecules, you have first to break the hydrogen bonds between them. The separation of molecules contributes to the viscous flow; the finding of a new partner for the hydrogen bond does not. 

The irreversible processes turn out \cite{bu2018a,bu2018b} to be responsible for the spectrum of of the dynamic heat capacity
\begin{equation}\label{pt}
	l_{irrev}(v)=\frac{1}{3\sqrt{2\pi}}\exp(v^2)\left(\ln(4\sqrt{2})-v\right)^{3/2},
\end{equation}
a slightly broadened Debye process around the relaxation time 1.6 $\tau_c$, a factor of thirteen longer than the Maxwell time. The irreversible process spectrum describes dynamic heat capacity data not only in a metallic glass, but also in three hydrogen bonding substances, the vacuum pump oil PPE, glycerol and propylene glycol \cite{bu2018a,bu2018b}, with $\tau_c$ determined from shear relaxation data of the same substances.

\begin{table}[htbp]
	\centering
		\begin{tabular}{|c|c|c|c|c|c|c|c|}
\hline
subst.                    &$T$    &$G$      &$\eta$&$\beta$ &  $f_K$  &$f_{exc}$  &$GJ_0$      \\
\hline
                          & K     &GPa      & GPas  &       &         &           &            \\
\hline
glycerol                  & 196   & 3.93    & 1.80 & 0.70   &  5.26   &  0.018    &   6.97     \\
propylene carbonate       & 159   & 1.49    & 0.523& 0.57   &  1.56   &  0.012    &   3.30     \\
PPE                       & 250   & 1.03    & 0.725& 0.52   &  1.0    &  0        &   2.64     \\
propylene glycol          & 180   & 3.78    & 0.17 & 0.70   &  3.67   &  0.02     &   5.19     \\
\hline
		\end{tabular}
	\caption{Parameters for the theoretical description of shear relaxation data (references see text) in the four hydrogen-bonded glass formers (PPE=5-polyphenylene ether).}
	\label{tab:rse1}
\end{table}

As pointed out in the theoretical paper \cite{bu2018b}, the simultaneous knowledge of irreversible and reversible relaxation processes from the shear data implies the knowledge of all Eshelby shear relaxation processes of the substance, and enables one to judge what one sees in other relaxation techniques. The application of this concept to dielectric and adiabatic compressibility data in non-hydrogen-bonded glass formers \cite{bu2018b} revealed that the scheme works very well, but the dielectric and compressibility signals required the multiplication of the total spectrum with $\exp(-\tau/\tau_D)$, where $\tau_D$ is a terminal time shorter than $\tau_c$, showing that the dielectric polarizability and the adiabatic compressibility equilibrate earlier than the terminal shear relaxation time
\begin{equation}\label{le}
	l(v)=(8f_Kl_{irrev}(v)+l_K(v))\exp(-\tau_c\exp(v)/\tau_D).
\end{equation}

In the application to dielectric spectra of hydrogen bonded substances, one finds that this simple scheme alone does not work. One must add a hydrogen-bond-correlation Debye decay at the time $\tau_D$ 
\begin{equation}\label{ltot}
	l_{tot}(v)=l_D\delta(v-\ln(\tau_D/\tau_c))+l_0l(v),
\end{equation}
where $l_0$ normalizes $l(v)$ to $1-l_D$.  Fig. 1 (b) shows that one can describe the dielectric spectrum of glycerol \cite{roessler} with this recipe.

$\tau_D/\tau_c$, and $l_D$ for the dielectric data are listed in Table II. In all four cases, one needs $l_D>0$ for a good fit. Also, one finds $\tau_D>\tau_c$ in propylene carbonate and propylene glycol. The findings demonstrate that the terminal Debye peak does not only appear in the monoalcohols \cite{mono}. This clear information is only accessible by the comparison to the shear spectra. For all four substances, the $\tau_D$-values are in the neighborhood of $\tau_c$, showing that the dipole decay coincides more or less with the lifetime of the Eshelby regions, unlike the monoalcohols \cite{mono}, where the Debye process occurs at much longer times. But note that in both cases the Debye process is not a different physical process. It is the combined result of many reversible and irreversible Eshelby transitions, and its Debye character is due to motional narrowing. The same is true for the normal mode of a polymer \cite{sokolov}.  

\begin{table}[htbp]
	\centering
		\begin{tabular}{|c|c|c|c|c|}
\hline
subst.                    &$T$    & $\Delta\chi$    & $\tau_D/\tau_c$ & $l_D$   \\
\hline
                          & K     &                 &                 &         \\
\hline
glycerol                  & 196   &     63.1        & 0.44            &  0.25   \\
propylene carbonate       & 159   &    100.0        & 1.53            &  0.43   \\
PPE                       & 250   &     1.89        & 0.82            &  0.13   \\
propylene glycol          & 180   &     65.0        & 3.1             &  0.42   \\
\hline
		\end{tabular}
	\caption{Parameters for the theoretical description of dielectric relaxation data (references see text) in the four hydrogen-bonded glass formers.}
	\label{tab:rse2}
\end{table}

Fig. 1 (b) shows also, as a dashed line, the calculated dielectric signal without the irreversible processes, which happens to describe the shifted depolarized dynamic light scattering data \cite{gabriel} very well. However, as argued by Thomas Blochowicz (see Supplemental Material), this agreement can be shown to be fortuitous.

Glycerol with its three strong oxygen hydrogen bonds per molecule has a much higher shear modulus than propylene carbonate, where the oxygen atoms do not have a hydrogen atom of their own, but link to the hydrogens bonded to carbon atoms. As a consequence, the deviation of the parameter $f_{K}$ from 1 in Table I of propylene carbonate is about a factor of ten smaller than the one of glycerol, and its shear modulus is not much higher than the 1 GPa of van-der-Waals bonded molecular glass formers \cite{bu2018b}. But otherwise, the results of the same analysis of propylene carbonate data \cite{pcnew} shown in Fig. 2 (a) and (b) and tabulated in Table I and II are very similar.

\begin{figure}[t]
\hspace{-0cm} \vspace{0cm} \epsfig{file=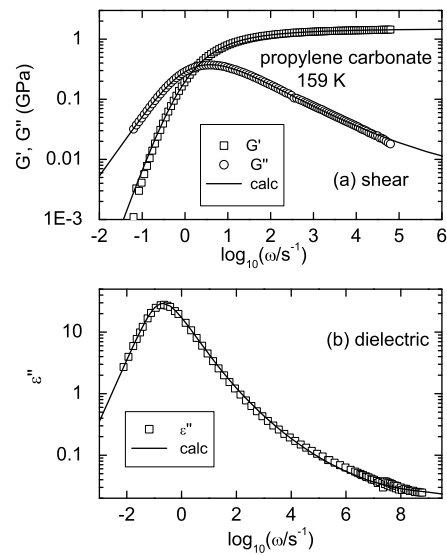,width=6 cm,angle=0} \vspace{0cm} \caption{(a) Measurement \cite{pcnew} of $G(\omega)$ in propylene carbonate at 159 K, continuous lines calculated with the parameters in Table I (b) Fit of dielectric data \cite{pcnew} at the same temperature in the same cryostat with the parameters in Table II.}
\end{figure}

PPE, where each oxygen bonds two phenyl rings, has even weaker hydrogen bonds, so weak that one gets a perfect fit of the shear data \cite{ppe} in terms of the original model with only the three parameters $G$, $\eta$ and $\beta$ in Table I. But the fit of the dielectric data improves markedly with the Debye parameter of 0.13 of Table II. The same fit in non-hydrogen bonded molecular glass formers supplies the Debye parameter zero within experimental error. The excess wing, not described by the fit, is nearly as strong as in the other three substances. This is demonstrated in Fig. 3 (b), showing that the dielectric signal is dominated by the strong hydrogen bond component which is not visible in the shear data.

\begin{figure}[b]
\hspace{-0cm} \vspace{0cm} \epsfig{file=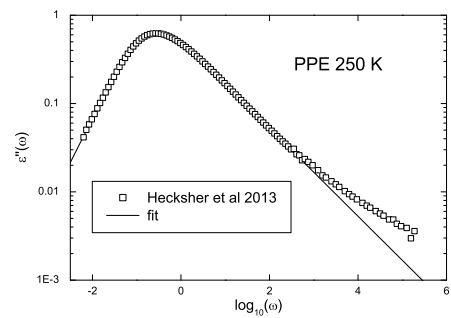,width=6 cm,angle=0} \vspace{0cm} \caption{Fit (continuous line, parameters Table II) of dielectric data in PPE at 250 K \cite{ppe}. Note the excess wing, which is not reproduced by the fit.}
\end{figure}

The last example, propylene glycol, has again a strong hydrogen bond component in its shear data \cite{ppg}. In the dielectric data \cite{ppgd}, one finds a large Debye parameter. 

Table I compiles the parameters for the four substances, including in the last row the total recoverable compliance $GJ_0$, which in glycerol is nearly a factor of three higher than in normal glass formers, showing once again that one needs much more recoverable processes to start the viscous flow, because of the influence of the hydrogen bonds. In all four cases, the shear analysis was done over the entire temperature range of the measurements, to look for a possible temperature dependence of the parameters. $f_K$ and $f_{exc}$ were found to be temperature-independent within experimental accuracy. For glycerol, the temperature-dependent shear moduli agree within a few percent with those of a transverse wave Brillouin scattering determination \cite{fio}, showing the high quality of both measurements.

These fits do not directly demonstrate the existence of a strong asymmetry of the double-well potentials. But additional evidence for strongly asymmetric double-well potentials is found in nonlinear dielectric data (see Supplemental Material).

To summarize, one can understand the shear relaxation of hydrogen-bonded undercooled liquids close to their glass transition in a recent theory of the highly viscous flow by taking the influence of hydrogen bond ruptures, well studied in water, into account. The hydrogen bond ruptures contribute to the shear fluctuations, but not to the viscous flow. They make the Kohlrausch tail double-well potentials strongly asymmetric and give rise to the excess wing, absent in non-hydrogen-bonded glass formers. One can describe shear, linear and non-linear dielectric, and dynamic specific heat data consistently within the theory.


\begin{thebibliography}{99}
\bibitem{cavagna} A. Cavagna, Phys. Rep. {\bf 476}, 51 (2009)
\bibitem{bb} L. Berthier and G. Biroli, Rev. Mod. Phys. {\bf 83}, 587 (2011)
\bibitem{royall} C. P. Royall and S. R. Williams, Phys. Rep. {\bf 560}, 1 (2015)
\bibitem{berthier} L. Berthier, J. Chem. Phys. {\bf 150}, 160902 (2019)
\bibitem{oh1} Th. Steiner, Angew. Chem. Int. Ed. {\bf 41}, 48 (2002) 
\bibitem{oh2} V. David, N. Grinberg, and S. C. Moldoveanu, in {\it Advances in Chromotography Vol. 54} (Eds: E. Gruschka, N. Grinberg), CRC, Boca Raton 2018, chap. 3 
\bibitem{mono} R. B\"ohmer, C. Gainaru, and R. Richert, Phys. Rep. {\bf 545}, 125 (2014)
\bibitem{sokolov} A. L. Agapov, V. N. Novikov, T. Hong, F. Fan, and A. P. Sokolov, Macromolecules {\bf 51}, 4874 (2018)
\bibitem{olsen} J. C. Dyre and N. B. Olsen, Phys. Rev. Lett. {\bf 91}, 155703 (2003)
\bibitem{gainaru} C. Gainaru, R. B\"ohmer, R. Kahlau, and E. R\"ossler, Phys. Rev. B {\bf 82}, 104205 (2010)
\bibitem{schneider} U. Schneider, R. Brand, P. Lunkenheimer, and A. Loidl, Phys. Rev. Lett. {\bf 84}, 5560 (2000)
\bibitem{gabriel} J. Ph. Gabriel, P. Zourchang, F. Pabst, A. Helbling, P. Weigl,  T. B\"ohmer, and Th. Blochowicz, Phys.Chem.Chem.Phys. {\bf 22}, 11644 (2020)
\bibitem{pabst} F. Pabst, J. Gabriel, T. B\"ohmer, P. Weigl, A. Helbling, T. Richter, P. Zourchang, Th. Walther, and Th. Blochowicz, J. Phys. Chem. Lett. {\bf 12}, 3685 (2021)
\bibitem{abs} U. Buchenau, G. D'Angelo, G. Carini, X. Liu, and M. A. Ramos, arXiv:2012.10139
\bibitem{bagchi} B. Bagchi, Chem. Rev. {\bf 105}, 3197 (2003)
\bibitem{laage} D. Laage and J. T. Hynes, Science {\bf 311}, 832 (2006)
\bibitem{bu2018a} U. Buchenau, J. Chem. Phys. {\bf 148}, 064502 (2018)
\bibitem{bu2018b} U. Buchenau, J. Chem. Phys. {\bf 149}, 044508 (2018)
\bibitem{plazek} D. J. Plazek, J. Phys. Chem. {\bf 69}, 3480 (1965)
\bibitem{tnb} D. J. Plazek and J. H. Magill, J. Chem. Phys. {\bf 45}, 3038 (1966)
\bibitem{bero} D. J. Plazek, C. A. Bero and I.-C. Chay, J. Non-Cryst. Solids {\bf 172-174}, 181 (1994)
\bibitem{roland} C. M. Roland, P. G. Santangelo, D. J. Plazek, and K. M. Bernatz, J. Chem. Phys. {\bf 111}, 9337 (1999) 
\bibitem{eshelby} J. D. Eshelby, Proc. Roy. Soc. {\bf A241}, 376 (1957)
\bibitem{falk} M. L. Falk and J. S. Langer, Phys. Rev. E {\bf 57}, 7192 (1998)
\bibitem{johnson} W. L. Johnson and K. Samwer, Phys. Rev. Lett. {\bf 95}, 195501 (2005)
\bibitem{bnap} R. B\"ohmer, K. L. Ngai, C. A. Angell and D. J. Plazek, J. Phys. Chem. {\bf 99}, 4201 (1993)
\bibitem{albena} A. I. Nielsen, T. Christensen, B. Jakobsen, K. Niss, N. B. Olsen, R. Richert, and J. C. Dyre, J. Chem. Phys. {\bf 130}, 154508 (2009)
\bibitem{glynew} M. H. Jensen, C. Gainaru, C. Alba-Simionesco, T. Hecksher, and K. Niss, Phys. Chem. Chem. Phys. {\bf  20}, 1716 (2018)
\bibitem{roessler} S. Adichtchev, T. Blochowicz, C. Tschirwitz, V. N. Novikov and E. A. R\"ossler, Phys. Rev. E {\bf 68}, 011504 (2003)
\bibitem{pcnew} C. Gainaru, T. Hecksher, N. B. Olsen, R. B\"ohmer, and J. C. Dyre, J. Chem. Phys. {\bf 137}, 064508 (2012)
\bibitem{ppe} T. Hecksher, N. B. Olsen, K. A. Nelson, J. C. Dyre and T. Christensen, J. Chem. Phys. {\bf 138}, 12A543 (2013)
\bibitem{ppg} C. Maggi, B. Jakobsen, T. Christensen, N. B. Olsen and J. C. Dyre, J. Phys. Chem. B {\bf 112}, 16320 (2008)
\bibitem{ppgd} C. Leon, K. L. Ngai, and C. M. Roland, J. Chem. Phys. {\bf 110}, 11585 (1999)
\bibitem{fio} F. Scarponi, L. Comez, D. Fioretto, and L. Palmieri, Phys. Rev. B {\bf 70}, 054203 (2004)
\end{thebibliography}
\end{document}



\title{Supplemental Material: Hydrogen bond dynamics at the glass transition}

\author{U. Buchenau}
 \email{buchenau-juelich@t-online.de}
\affiliation{%
Forschungszentrum J\"ulich GmbH, J\"ulich Centre for Neutron Science (JCNS-1) and Institute for Complex Systems (ICS-1),  52425 J\"ulich, GERMANY}%

\date{October 31, 2021}

\begin{abstract}
The argument of Thomas Blochowicz for the depolarized dynamical scattering is presented and discussed. The consequences of the main paper for the nonlinear dielectric effects are worked out.
\end{abstract}

\keywords{Glass transition; Hydrogen bonds}
\maketitle

\section{Depolarized dynamical light scattering: Blochowicz argument}

The imaginary peak of the depolarized dynamical light scattering data in glycerol lies a factor of three higher than the one of the dielectric data \cite{gabriel}. In an earlier draft \cite{hb2} of the present paper, this led to the hypothesis that one sees only reversible relaxations in the depolarized dynamical light scattering.

In a discussion with the Author, Thomas Blochowicz pointed out that one could check the hypothesis in DC704, where the two peaks happen to coincide \cite{pabst}.

This is done in Fig. 1, which shows the dielectric spectrum \cite{tina} of DC704 and its fitted \cite{ac} full and reversible parts, with a reversible part peak frequency a factor of 2.5 higher than the one of the full spectrum.

\begin{figure}[t]
\hspace{-0cm} \vspace{0cm} \epsfig{file=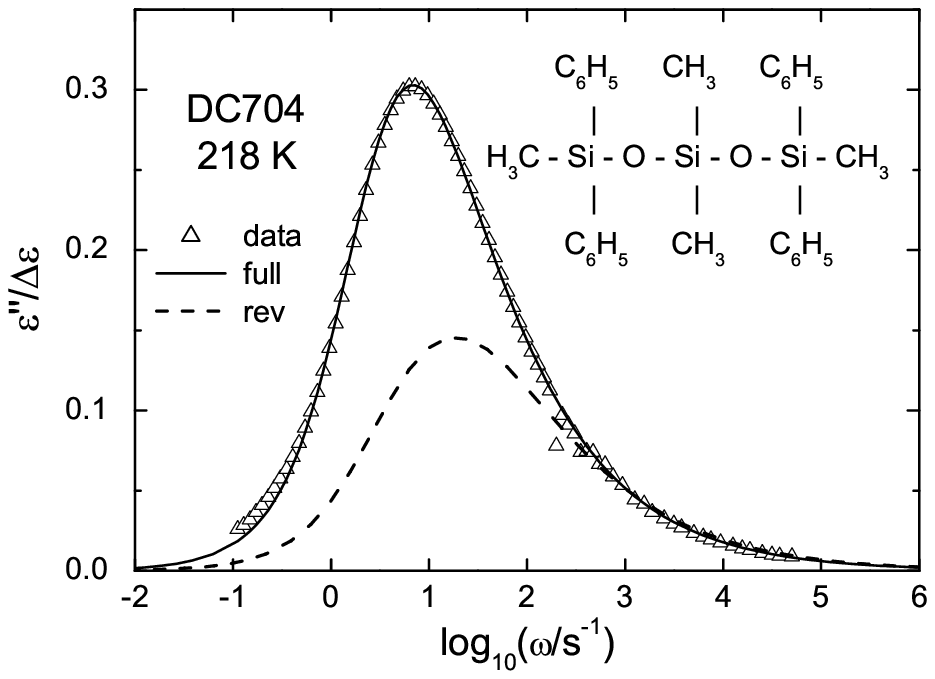,width=6 cm,angle=0} \vspace{0cm} \caption{Dielectric data \cite{tina} in DC704 at 218 K, together with their theoretical fit \cite{ac} (continuous line). Taking only the reversible processes to be responsible for the depolarized dynamical light scattering, one predicts the dashed line with a peak shift which is not observed in experiment \cite{pabst}. The structure of DC704 with its two Si-O-Si bonds in the center is shown in the upper right side of the figure.}
\end{figure}

The comparison to the measured depolarized dynamical light scattering data of DC704 \cite{pabst} shows clearly that the former hypothesis of the Author is wrong. In fact, the hypothesis has the additional weakness that, according to theory \cite{bu2018b}, at the crossover from reversible to irreversible structural transitions their nature does not change, so their visibility in the depolarized dynamic light scattering should not change.

Looking for an alternative explanation of the upwards peak shift from dielectrics to depolarized dynamical light scattering in glycerol, one remembers that for the simple case of isotropic rotational diffusion of a molecular dipole Debye \cite{debye} predicts the peak in $\epsilon''$ at $\omega=2D_r$ ($D_r$ rotational diffusion constant) and Berne and Pecora \cite{pecora} predict the imaginary peak in the depolarized dynamical light scattering at $\omega=6D_r$, a factor of three higher.

Naturally, the molecular motion in the $\alpha$-relaxation of glycerol is not a simple rotational diffusion. On the other hand, the terminal stage of the motion at the time $\tau_D$, where the many Eshelby transitions in which the molecule participated have removed its initial connection to its neighbors, is essentially a small-angle motion according to NMR evidence \cite{bohmer}. So the measured peak shift in glycerol \cite{gabriel} seems understandable, even quantitatively.

But this poses the question why the peak shift is not observed in DC704 \cite{pabst}. 

To understand this, look at the structure of the DC704 molecule in Fig. 1, with its two very flexible Si-O-Si bond bending degrees of freedom. The dipole moment of each of them is given by the displacement of the oxygen atom from the line connecting the two silicon atoms. One does not expect the very large DC704 molecule to rotate much during the whole $\alpha$-process, but the Si-O-Si dipole moments will change in every Eshelby transition in which the molecule participates. If one idealizes each Si-O-Si bond as a dipole making large jumps in a fixed plane perpendicular to the line connecting the two silicon atoms, with a 180-degree jump equally probable as a 90-degree jump, one does indeed predict equal peak positions in dielectrics and depolarized dynamical light scattering.  

\section{Nonlinear dielectric effects} 

The first two decades of this century have provided high quality experimental nonlinear dielectric results, all of them taken close to the glass transition in hydrogen-bonded undercooled liquids \cite{weinstein,weinstein2,wang,huanga,huang,brun,tca,brun2,brun3,bauer1,bauer3,bauer2015,samanta,r2016,chi5}. These data hold the promise for a deep insight into the nature of the highly viscous flow - if one could understand them. 

Previous attempts \cite{brun2,bu2017} to explain the experimental findings in terms of the nonlinear dielectric response of asymmetric double-well potentials \cite{brun3,gregor} were not very successful, but Ranko Richert's phenomenological model \cite{weinstein,wang,huanga,huang,tca,r2016}, sometimes also called box model, is impressively able to explain all nonlinear dielectric findings without free parameters. The model requires the hole-burning assumption \cite{schiener} that the relaxations heat up into a state corresponding to a higher temperature when they absorb the electric energy.

The results of the present paper enable a new interpretation of the Richert model in terms of asymmetric double-well potentials, attributing the change of the relaxation in a strong alternating electric field to a reduction of the asymmetry $\Delta$ of the double-well potential.

Experimentally \cite{huang}, one finds that the energy absorption plays a large role in the nonlinear change of $\epsilon''(\omega)$, but does practically not influence the nonlinear signal at $3\omega$. Therefore one has to begin with an analysis of the average asymmetry $\Delta$ and the dipole moment change $M$ of the Eshelby transitions from the third order nonlinear susceptibility $\chi_3^{(3)}(\omega)$, expressed in a convenient dimensionless form \cite{brun2}
\begin{equation}\label{x3}
	X_3(\omega)=\frac{NkT}{\epsilon_0\Delta\chi^2}\left|\chi_3^{(3)}(\omega)\right|.
\end{equation}
Here $N$ is the number density of molecules, and $\Delta\chi$ is the susceptibility difference between low and high frequencies.

For a constant asymmetry density, and weighting the asymmetries with the prefactor $1/\cosh^2(\Delta/2k_BT)$ (the weakening factor for the relaxation strength of an asymmetric double well potential), one expects the average value $\Delta/k_BT=1.317$, the one which fulfills the condition $\tanh^2(\Delta/2k_BT)=1/3$ \cite{gregor} for a hump in $X_3(\omega)$. This is supposed to be the case of non-hydrogen bonded glass formers \cite{bu2018b}. For this case, one finds the Gilroy-Phillips model \cite{gilroy} fulfilled. The validity of this model in the glass phase close to the glass transition has been recently demonstrated for several non-hydrogen bonded glass formers \cite{abs}. 

For the constant asymmetry density, the average weakening factor by the asymmetry $1/\cosh^2(\Delta/2k_BT)=2/3$. The enhancement factors $f_K$ between three and five for glycerol and propylene glycol, where the hydrogen bonds determine the shear modulus, indicate a weakening factor from the asymmetry between 2/9 and 1/9, implying an average asymmetry of the double-well potentials around $\tau_c$ of about 2 $k_BT$.

\begin{figure}[t]
\hspace{-0cm} \vspace{0cm} \epsfig{file=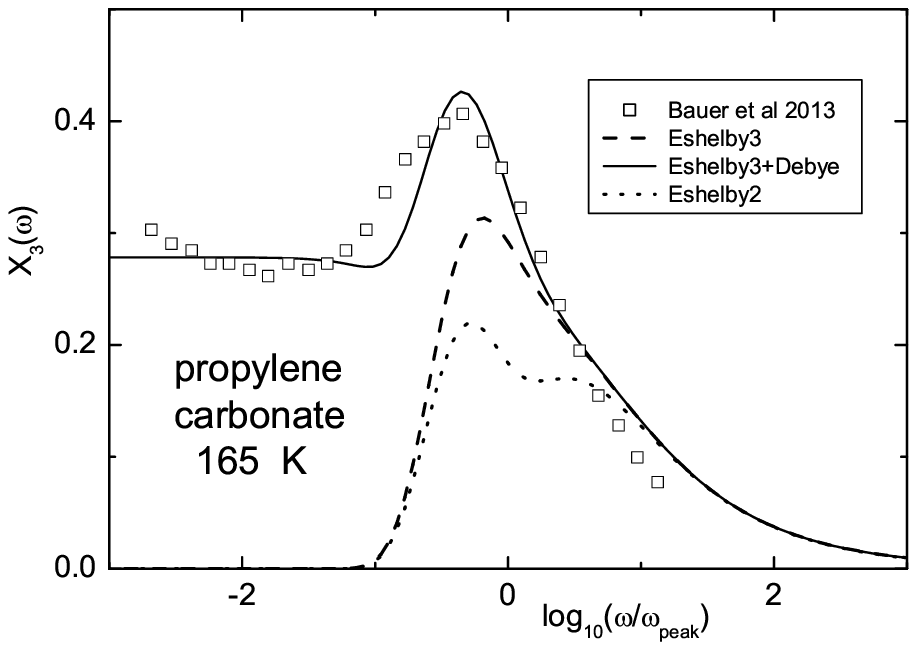,width=6 cm,angle=0} \vspace{0cm} \caption{Measurement of the hump in the nonlinear dielectric effect at $3\omega$ in propylene carbonate at 165 K \cite{bauer3}, compared to theoretical fits with the average asymmetry $\Delta=2k_BT_g$ (Eshelby2) and $\Delta=3k_BT_g$ (Eshelby3). Parameters of the fit with $\Delta=3k_BT_g$ listed in Table I.}
\end{figure}

But it turns out that the data for $X_3(\omega)$ exclude the value 2 $k_BT$. Fig. 2 shows a measurement of $X_3(\omega)$ in propylene carbonate at 165 K \cite{bauer3}, together with two fits in terms of the theoretical barrier density of the main paper, one for $\Delta=2k_BT$ (the dotted Eshelby2-curve) and the other for $\Delta=3k_BT$ (the dashed Eshelby3-curve). The first one is still too close to the hump condition $\tanh^2(\Delta/2k_BT)=1/3$, where one has strong frequency oscillations in the nonlinear signal \cite{gregor}. The pronounced shoulder in the calculated curve is due to the interference effects from neighboring relaxations. At $\Delta=3k_BT$, the oscillations are much less pronounced and lead only to a weak shoulder, much more consistent with the measured data. 

Both fits required the assumption of a negligible nonlinear signal from the excess wing, in order to get the right slope at high frequencies. At $3\omega$, one has a low-frequency cutoff $\exp(-1/3\omega\tau_{dip})$, with $\tau_{dip}=1.6\tau_c\tau_D/(1.6\tau_c+\tau_D)$, both from the decay of the dipole moment and from the decay of the Eshelby regions.

For the fit of the whole curve (continuous line in Fig. 2) one has to add the third order effect \cite{coffey} from the terminal Debye relaxation at $\tau_D$. For completely uncoupled molecular dipole moments \cite{bu2017}, $X_3(0)=f_{On}/5$. But for real glass formers, one has to introduce a scaling parameter $X_3(0)=f_3f_{On}/5$, in this case $f_3=0.5$.

The fit with $\Delta=3k_BT$ required a dipole moment change $M/\mu=2.39$ in the Eshelby transitions, where $\mu=5.67$ D is the dipole moment calculated from $\Delta\chi$ and the density at 166 K \cite{tca}, a Kirkwood factor 1.16 larger than the molecular dipole moment 4.9 D of the isolated molecule. The fit further required the validity of the asymmetric double-well equations \cite{gregor} not only for the reversible, but also for the irreversible Eshelby transitions.

\begin{figure}[t]
\hspace{-0cm} \vspace{0cm} \epsfig{file=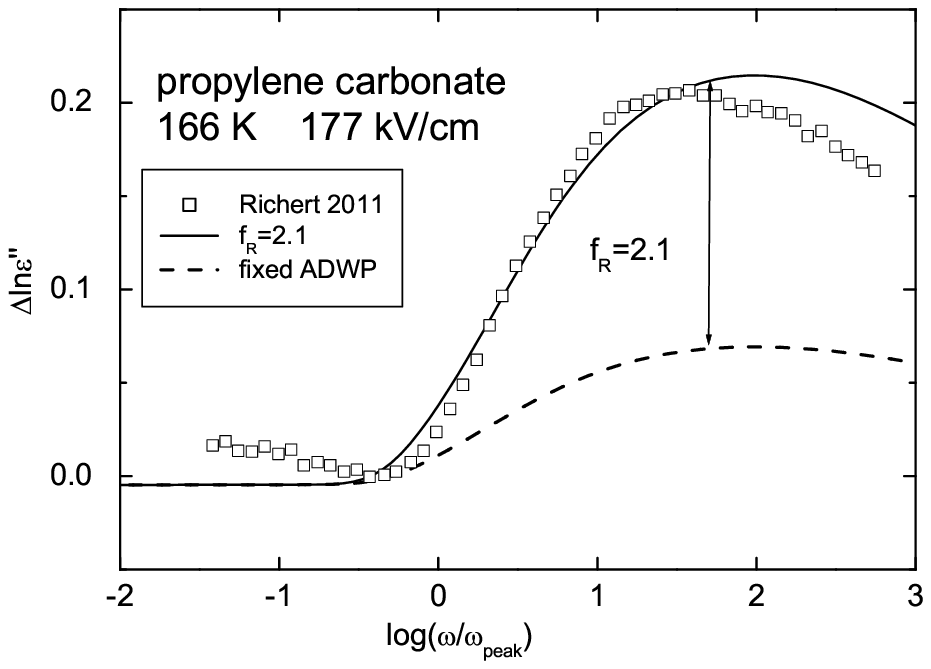,width=6 cm,angle=0} \vspace{0cm} \caption{Fit of measured nonlinear $\Delta\epsilon''$-data \cite{tca} in propylene carbonate in terms of a Richert factor of 2.1 and an average asymmetry $3k_BT_g$.}
\end{figure}

Knowing $\Delta$ and $M$ from the fit of $X_3(\omega)$, one is able to calculate (with the lifetime limitation $\exp(-1/\omega\tau_{dip})$) the nonlinear change $\Delta\epsilon''$ of the imaginary part of $\epsilon''(\omega)$ under the influence of a strong alternating electric field for fixed asymmetric double-well potentials \cite{gregor}. This, together with a small and essentially constant negative contribution from the Debye relaxation at low frequency, leads to the dot-dashed line in Fig. 3, which at higher frequencies turns out to be an average factor of 3.1 lower than measured propylene carbonate data \cite{tca}. 

In the spirit of the Richert model, we attribute this enhancement to a change of the relaxation under the influence of the strong alternating field. But instead of postulating a transformation of the relaxation into one at higher temperature, we assume a decrease of the asymmetry $\Delta$. For an asymmetry $3k_BT$, one only needs a decrease by less than 0.2 $k_BT$ to achieve the measured \cite{tca} enhancement in an alternating field of 177 kV/cm.

We quantify this Richert enhancement $1+f_R$ by the Richert factor $f_R$, in the case of Fig. 3 $f_R=2.1$. Its microscopic origin is obvious: The strong alternating electric field forces a higher occupation of the upper well, thus giving it a longer time to relax its high energy.

In order to understand the process, imagine a double-well potential which has the elastic misfit zero in its lower well, and a strong elastic misfit leading to $\Delta=3k_BT$ in the other. Whenever the system stays in the upper well, the energy $\Delta$ will decrease by the elastic shear relaxation outside the region, even if the region itself keeps its structure unchanged.

\begin{table}[htbp]
	\centering
		\begin{tabular}{|c|c|c|c|c|c|c|c|c|}
\hline
subst.  &$T$    & $\Delta\chi$    &$\mu$   &$M/\mu$    & $E_e$     & $f_R$ &$f_{R,th}$ &$f_e$  \\
\hline			
        & K     &                 &  D     &           &  kV/cm    &       &           &       \\
\hline
PC      & 166   &    130.0        & 5.67   & 2.39      &  177      &  2.1  &   2.0     &  0.5  \\
glyc    & 204.7 &     62.7        & 4.26   & 2.07      &  671      &  1.0  &   1.2     &  0.3  \\
\hline
		\end{tabular}
	\caption{Parameters for the theoretical description of nonlinear dielectric relaxation data (references see text) in the two hydrogen-bonded glass formers PC=propylene carbonate and glyc=glycerol for an assumed average asymmetry $\Delta=3k_BT$. The theoretical Richert factors $f_{R,th}$ were calculated from equs. (\ref{rich}) and (\ref{greg}) with $\beta=0.57$ for propylene carbonate and $\beta=0.7$ for glycerol, the values in Table I of the main paper.}
	\label{tab:rse2}
\end{table}

As long as the change of the shear misfit remains small, the energy of the lower well stays practically the same, because, unlike the energy of the upper well, it increases only with the square of the small change. Therefore the changes are not reversed in the times which the system spends in the lower well, but sum up within the times spent in the upper well over the lifetime of the inner region.

For a shear relaxation following a Kohlrausch law with exponent $\beta$, this leads to a small difference between its asymmetry $\Delta(0)$ in the moment of its creation and the average value $\Delta$ during its lifetime
\begin{equation}
	\Delta(0)-\Delta=\Delta\left(\frac{1}{2\exp(\Delta/k_BT)+2}\right)^\beta,
\end{equation}
a relatively small effect.

But this small effect becomes observable in a strong alternating alternating electric field $E$, because the field enhances the occupation of the upper well by the factor $1+a^2/2$, where $a=f_{On}EM/k_BT$. It is straightforward to calculate the nonlinear increase $\Delta\epsilon''/\epsilon''$ from the additional lowering of the asymmetry
\begin{equation}\label{rich}
	\frac{\Delta\epsilon''}{\epsilon''}=\left(\frac{1}{2\exp(\Delta/k_BT)+2}\right)^\beta\frac{\Delta\delta^2a^2}{2},
\end{equation}
where $\delta=\tanh(\Delta/2k_BT)$.

\begin{figure}[t]
\hspace{-0cm} \vspace{0cm} \epsfig{file=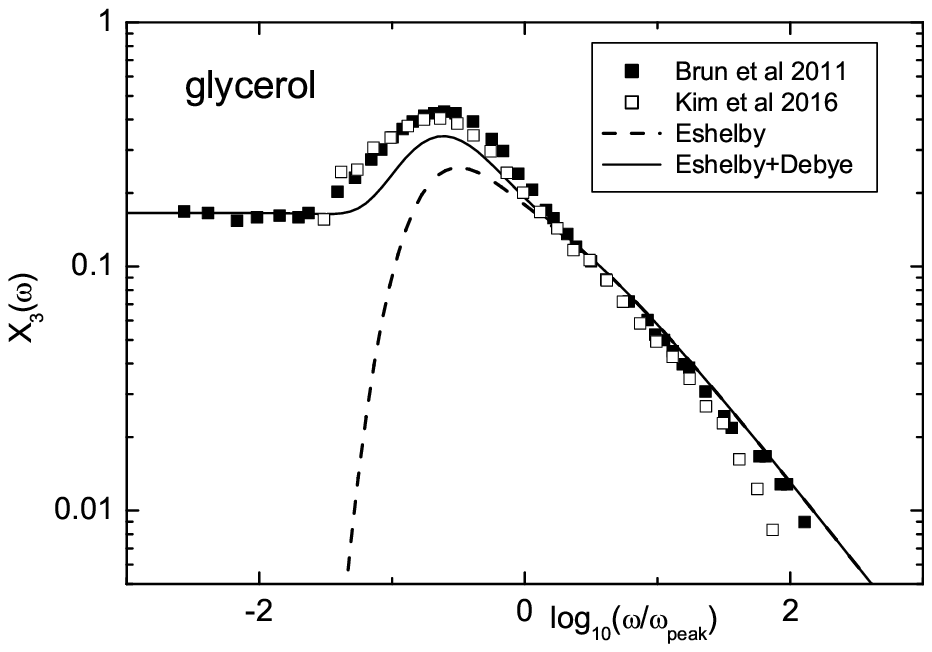,width=6 cm,angle=0} \vspace{0cm} \caption{Measurements of the hump in the nonlinear dielectric effect at $3\omega$ in glycerol at 204.7 K \cite{brun2} (full squares) and at 210 K \cite{r2016} (open squares), compared to the theoretical fit with $\Delta=3k_BT_g$ (parameters of the fit listed in Table I).}
\end{figure}

\begin{figure}[b]
\hspace{-0cm} \vspace{0cm} \epsfig{file=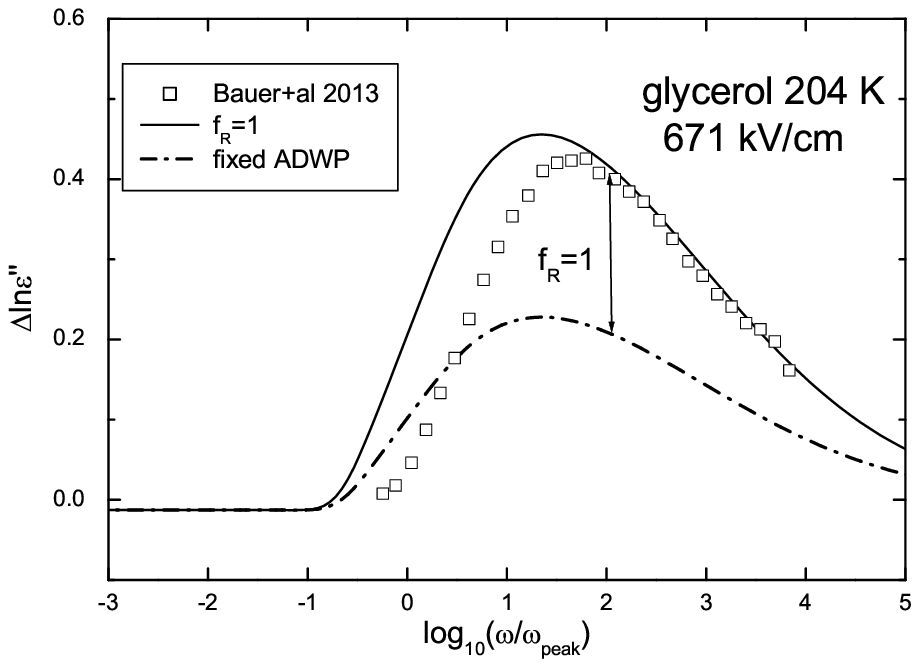,width=6 cm,angle=0} \vspace{0cm} \caption{Fit of measured nonlinear $\Delta\epsilon''$-data \cite{bauer1} in glycerol in terms of a Richert factor of 1.0 and an average asymmetry $3k_BT_g$.}
\end{figure}

This has to be compared to the nonlinear effect in the fixed potential \cite{gregor} at $\omega\tau=1$
\begin{equation}\label{greg}
	\frac{\Delta\epsilon''}{\epsilon''}=(3\delta^2/5-1/4)\frac{3a^2}{10}
\end{equation}

The ratio of the quantity calculated in eq. (\ref{rich}) to the one from eq. (\ref{greg}) supplies the Richert factor for a given $\Delta$ and $\beta$. Table I shows that the theoretical Richert factor $f_{R,th}=2.0$ determined in this way (with $\Delta=3k_BT$ and the value $\beta=0.57$ from Table I of the main paper) agrees within small error bars with the fitted value 2.1.

Note that this explanation of the Richert effect does not contradict the validity of the fixed potential assumption for $X_3(\omega)$. The effect is only strong in the $\omega$-signal and has little influence at $3\omega$, consistent with the observed absence of long-time changes \cite{huang} at $3\omega$ after switching on a strong alternating electric field. 

Figs. 4 and 5 show the similar results of the same procedure for glycerol. In this case, the Richert factor equals 1.0, again in good agreement with the theoretical value 1.2 calculated with $\Delta=3k_BT$ and $\beta=0.7$ from Table I of the main paper. The calculation shows that the different $f_R$-values in propylene carbonate and glycerol are due to a difference in the Kohlrausch $\beta$-values.

\begin{figure}[t]
\hspace{-0cm} \vspace{0cm} \epsfig{file=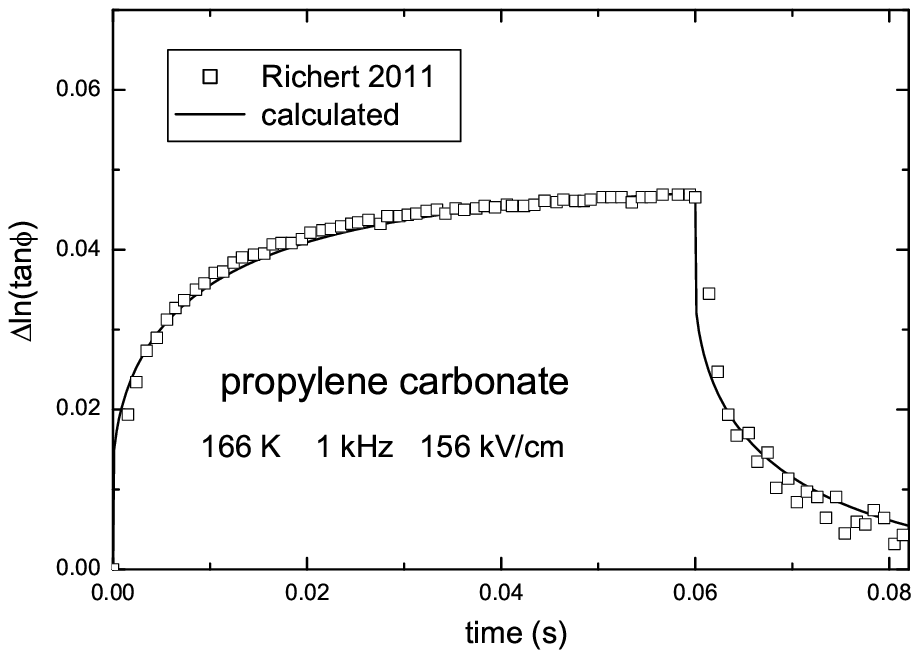,width=6 cm,angle=0} \vspace{0cm} \caption{Measured nonlinear on-switch and off-switch  $\Delta\ln(\tan{\Phi})$-data for a strong alternating electric field in propylene carbonate \cite{tca}, described in terms of the parameters in Table I.}
\end{figure}

The Kirkwood factor of about 1.6 of glycerol is markedly higher than the one of 1.16 of propylene carbonate, showing a higher positive correlation of the molecular dipole moments. This is also seen in the $f_3$-values, which determine the Piekara factor at zero frequency. They reduce the expected Langevin decrease for noninteracting dipoles by a factor of 2 in propylene carbonate, and a factor 3.3 in glycerol, a markedly stronger effect than the Kirkwood factor of the linear response.

In this context, it is interesting to note that this effect is even stronger in the fifth order nonlinear measurements \cite{chi5}, with $\chi_5^{(5)}(0)$-values of 9 for propylene carbonate and 2.5 for glycerol in units of 10$^{-33}$ m$^4$/V$^4$. Translating this into $X_5(0)$ with its theoretical free dipole value $2f_{On}^2/35$ ( not $2f_{On}^2/7$ as stated erroneously in reference \cite{bu2017}), one finds that one needs a reduction by more than a factor of ten in both cases.   

Fig. 6 shows that the propylene carbonate parameters in Table I are able to describe time-dependent on-switching and off-switching experiments of the strong alternating field \cite{tca}, merely assuming that the fixed potential component relaxes exponentially with the relaxation time of the potential, while the Richert component relaxes with the Kohlrausch function $\exp(-(t/1.6\tau_c)^\beta)$. The value of $\tau_c=4.02$ ms is taken from fits of the shear relaxation data \cite{pcnew}.  

In the off-switch, one has to take into account that the Kohlrausch function is not fully saturated (in the case of Fig. 6 only to 94 percent). One can take the effect approximately into account by replacing $\exp(-(t/1.6\tau_c)^\beta)$ for the off-switch by $(\exp(-(t/1.6\tau_c)^\beta)-0.06)/0.94$, which describes the terminal data in Fig. 6 reasonably well.

To conclude, the hydrogen bond scheme of the main paper is able to supply a quantitative description of nonlinear dielectric data in terms of double-well potentials with an average asymmetry of about $3k_BT_g$, somewhat lower than the value $5k_BT_g$ measured at the excess wing \cite{gainaru} and close to the value $3.8BT_g$ measured at the secondary relaxation peak of tripropylene glycol \cite{olsen}. Reversible and irreversible Esdhelby transitions contribute equally, possibly because in the detailed balance only the asymmetry counts, and it does not matter how one comes from one state to another, whether by direct transition or whether through an intermediate state.